\documentclass[a4paper,11pt]{article}
\usepackage{jinstpub} 
\usepackage{lineno}
\usepackage{siunitx}

\title{\boldmath Optimizing Dynamic Aperture Studies with Active Learning}

\author[1]{D. Di Croce,}
\author[2]{M. Giovannozzi,}
\author[3]{E. Krymova,}
\author[1]{T. Pieloni,}
\author[2]{S. Redaelli,}
\author[1,4]{M. Seidel,}
\author[2]{R.~Tomás,}
\author[2]{F. F. Van der Veken}
\affiliation[1]{École Polytechnique Fédérale de Lausanne, Lausanne, Switzerland}
\affiliation[2]{CERN, Geneva, Switzerland}
\affiliation[3]{The Swiss Data Science Center, Zurich, Switzerland}
\affiliation[4]{Paul Scherrer Institut, Villigen, Switzerland}

\emailAdd{davide.dicroce@epfl.ch}

\abstract{
Dynamic aperture is an important concept for the study of non-linear beam dynamics in circular accelerators. It describes the extent of the phase-space region where a particle's motion remains bounded over a given number of turns. Understanding the features of dynamic aperture is crucial for the design and operation of such accelerators, as it provides insights into nonlinear effects and the possibility of optimising beam lifetime. The standard approach to calculate the dynamic aperture requires numerical simulations of several initial conditions densely distributed in phase space for a sufficient number of turns to probe the time scale corresponding to machine operations. This process is very computationally intensive and practically outside the range of today's computers. In our study, we introduced a novel method to estimate dynamic aperture rapidly and accurately by utilising a Deep Neural Network model. This model was trained with simulated tracking data from the CERN Large Hadron Collider and takes into account variations in accelerator parameters such as betatron tune, chromaticity, and the strength of the Landau octupoles. To enhance its performance, we integrate the model into an innovative Active Learning framework. This framework not only enables retraining and updating of the computed model, but also facilitates efficient data generation through smart sampling. Since chaotic motion cannot be predicted, traditional tracking simulations are incorporated into the Active Learning framework to deal with the chaotic nature of some initial conditions. The results demonstrate that the use of the Active Learning framework allows faster scanning of the configuration parameters without compromising the accuracy of the dynamic aperture estimates.
}


\begin{document}
\maketitle
\flushbottom

\section{Introduction}
The exploration of dynamic aperture (DA), which is defined as the extent of the connected phase-space region within which the dynamics of a single particle remains bounded in circular accelerators (see, e.g. Ref.~\cite{PhysRevE.53.4067} and references therein), offers invaluable insights into the non-linear beam dynamics of non-interacting particles and the mechanisms contributing to beam losses~\cite{da_and_losses}. This knowledge is crucial to optimise machine performance.

To calculate the DA numerically, one must track a multitude of initial conditions in phase space over numerous turns (at least $10^5$ turns). This method imposes significant computational demands, particularly in the case of large accelerator rings such as the CERN Large Hadron Collider (LHC)~\cite{LHCDR}, its luminosity upgrade HL-LHC~\cite{TDR}, or the Future Circular Colliders under study at CERN~\cite{FCC-eeCDR,FCC-hhCDR}. The difficulty is not in the number of initial conditions for which parallelisation can be applied, but rather in the number of turns, as no optimisation can be applied to reduce the total time required to obtain the simulation results. For these reasons, intense efforts have been devoted to the determination of analytical scaling laws that provide the evolution of the DA over time. In this way, numerical simulations can be used to successfully determine the evolution of DA well beyond the maximum simulated number of turns~\cite{PhysRevE.57.3432,Bazzani:2019csk}.

In recent years, machine learning (ML)-based approaches have been explored to compute efficient surrogate models of DA (e.g., see~\cite{schenk:ipac21-tupab216,casanova:2023}). An alternative approach has emerged that takes advantage of deep learning (DL) to quickly and accurately predict DA for machine configurations that are unknown to the model~\cite{dicroce_ipac23}. This was achieved by training a Deep Neural Network (DNN) on a substantial data set of simulated initial conditions. This DNN effectively captures the intricate relationship among accelerator parameters, initial conditions, and the resulting DA values. In this study, we take this integration of ML techniques a step further by incorporating an Active Learning (AL) framework. To this end, we introduce an error estimator in conjunction with the DA model to assess the uncertainty in the predictions. This addition allows the AL algorithm to conduct intelligent sampling. In other words, it can prioritise which new accelerator configurations to simulate first, guided by the magnitude of the predicted DA error. This approach aims to facilitate the rapid estimation of DA and its associated error for new accelerator parameters while simultaneously expanding the initial data set in an efficient manner, all with the added objective of enhancing the performance of the ML model.

\section{Simulated Samples}
\subsection{Simulating Accelerator Configurations}

To train the DNN, we simulated several accelerator configurations using MAD-X~\cite{madx} and the 2023 LHC lattice at injection energy (\SI{450}{GeV}) ~\cite{lhcoptics2023}. We varied five accelerator parameters, namely the betatron tunes $Q_x, Q_y$, chromaticities $Q'_x, Q'_y$, strength of the Landau octupoles (using the current $I_\mathrm{MO}$) and the realisations (also called seeds) of the magnetic field errors that have been assigned to the various magnet families. Furthermore, both Beam~1 (clockwise) and Beam~2 (counter-clockwise) have been considered in these studies. 
We performed a random uniform grid search of the following parameters: $Q_x \in [62.1,62.5]$ and $Q_y \in [60.1,60.5]$, with steps of size $5\times 10^{-3}$, $Q' \in [0,30]$, with steps of size $2$, $I_\mathrm{MO} \in [\SI{-40}{\ampere},\SI{40}{\ampere}]$, with step size $\SI{5}{\ampere}$, and 60 realisations of the magnetic errors for both beams. 
Furthermore, we trained the model on the 2016 and 2023 LHC optics for comparison; the main difference between them is the implementation of the Achromatic Telescopic Squeezing (ATS)~\cite{PhysRevSTAB.16.111002,PhysRevAccelBeams.24.021002} optics in the 2023 LHC configuration. 

\subsection{Simulating the DA}

To simulate DA, we probed the phase space by tracking the initial conditions with XSuite~\cite{xsuite_doc,xsuite_gianni} for $10^5$ turns. These particles are initially distributed in polar coordinates, allowing us to assess the stability limit for each angle, the so-called angular DA, i.e. the largest amplitude that survives for up to $10^5$ turns. 
With this approach, determining the stable limit is straightforward by finding the particle with the largest amplitude that remained bounded. 
However, it should be noted that this method may not be as precise for concave DA regions. 
Furthermore, the angular DA approach offers a higher density of initial conditions for smaller radii, which is particularly valuable when calculating loss rates and considering a Gaussian particle distribution. Unlike Cartesian coordinates, where the coordinates are evenly distributed in both x and y, polar coordinates provide a higher particle density at smaller radii. For this reason, probing with polar coordinates improves the precision of loss-rate measurements.

To speed up tracking, we performed an initial scan of a set of conditions uniformly distributed in 8 polar angles in $[0, \pi/2]$ and 33 radial amplitudes in $[0.0, 20 \sigma]$, to identify the value of the angular DA for a given angle, as well as the first amplitude where the particle does not survive more than $10^3$ turns (limit of the fast-loss zone). Then a finer scan of the initial conditions was performed in an amplitude range between the stable zone $-2 \sigma$ and the fast loss zone $+2 \sigma$, with radial steps of 0.06 $\sigma$ along 44 polar angles in $[0, \pi/2]$. 

\begin{figure}[!htbp]
\centering
\includegraphics[width=0.55\linewidth, height=0.5\linewidth]{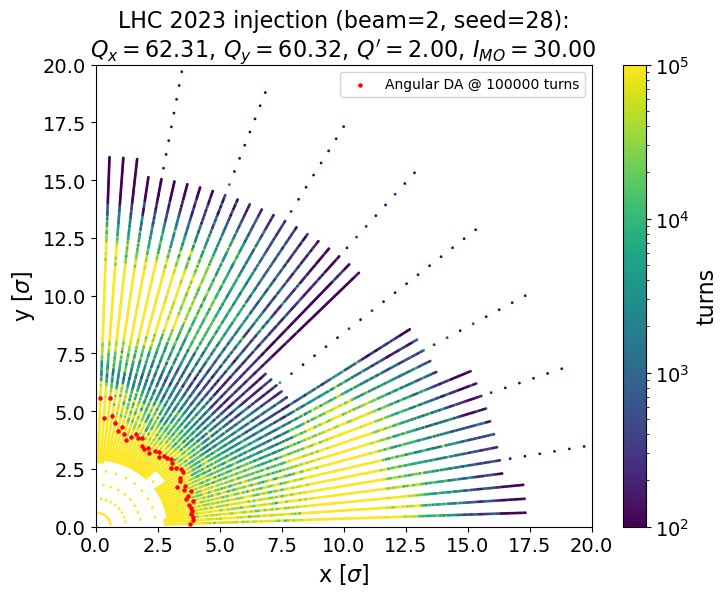}
\caption{Stability time for a distribution of initial conditions used to compute the DA for a specific accelerator configuration. The angular DA at $10^5$ turns is shown in red.}
\label{fig:input_DA}
\end{figure}

This procedure enables for a more detailed scan in phase space, specifically focusing on the region of interest. This approach not only enhances the precision of the angular DA but also accelerates tracking by an average factor of three compared to probing the whole phase space, as established empirically. An example of the results of these computations in the $x-y$ space is shown in Fig.~\ref{fig:input_DA} for a specific accelerator configuration. It is important to note that the accelerator configuration depicted in the graph does not utilise the operational tunes, which would have provided a DA of about 7 $\sigma$.

In addition, our objective includes understanding the evolution of beam losses across various accelerator configurations. Our approach involves studying how the DA changes while monitoring the survival of the particles at different numbers of turns. For this reason, the stability times have been binned using 12 distinct stability-time limits, namely 1, 5, 10, 20, 40, 50, 60, 70, 80, and 100 thousand turns. 

The number of samples used for our model is calculated by multiplying three components: the number of accelerator configurations present in the data set, the number of angles used to probe the phase space (44), and the number of bins (12) used to probe the evolution of the stability time, which gives a substantial 528-fold expansion in the data set size, significantly enhancing the likelihood of achieving convergence during model training, while also allowing us to monitor the shape of the stability limits and its evolution. From this sample size, 10\% of the samples were used for validation and 10\% to test the performance of the model. 

\subsection{Data preprocessing}

To gain more insight into beam dynamics, we added other variables in addition to those used to differentiate the accelerator configuration in the simulations, which are calculated by MAD-X and PTC~\cite{ptc_doc}. We considered seven anharmonicity coefficients, i.e. the amplitude detuning terms up to the second order, to take into account the impact of non-linear effects on the beam dynamics. Furthermore, we included the maximum value of the Twiss parameters $\alpha$ and $\beta$, and the phase advance ($\mu_{x,y}$) between IP1 and IP5 to take into account the impact on the optical parameters from the nonlinear field errors.

Moreover, we carried out standardisation on the continuous input variables, excluding beam and seed variables. This involved normalising the distributions of these variables using their respective averages and standard deviations. This crucial step guarantees uniform scaling across all input features, facilitating faster model convergence and model stability. As a result, this leads to improved model performance and interpretability. 
Furthermore, we restrict the angular DA values to a maximum of 18 $\sigma$. Values exceeding this threshold are considered outliers in a distribution ranging from 0 to 20 $\sigma$. This restriction is implemented to prevent the influence of extremely high angular DA values on the regressor, as discussed in~\cite{cook}.

\section{DA Regressor}
\subsection{Network Architecture}

Similarly to our previous study~\cite{dicroce_ipac23}, we used a Multilayer Perceptron (MLP)~\cite{lecun} as the architecture for our surrogate model. This MLP is responsible for examining accelerator parameters, the polar angle, and tracked turns to regress the angular DA.
It is essential to acknowledge that MLPs have some limitations; one of them is their lack of enforceability of constraints on particle survival time, which means that they cannot inherently ensure that angular DA decreases as we extend particle tracking over more turns for the same machine configuration. On the contrary, recurring neural network (RNN) architectures have the ability to address this issue, but require careful convergence~\cite{rnn_gradient}, making MLPs an attractive choice when we prioritise faster and more straightforward model training.
Another disadvantage of MLPs is their lack of geometry constraints, which means that the prediction of the angular DA is independent of neighbouring values. In contrast, image-based networks can use neighbouring information~\cite{imagenet}. However, they require additional simulated sets of accelerator configurations to achieve convergence. 

\begin{figure}[htbp]
\centering
\includegraphics[height=0.48\linewidth]{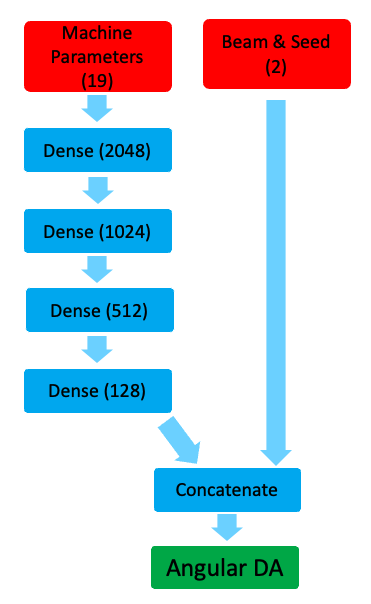}
\qquad
\includegraphics[height=0.48\linewidth]{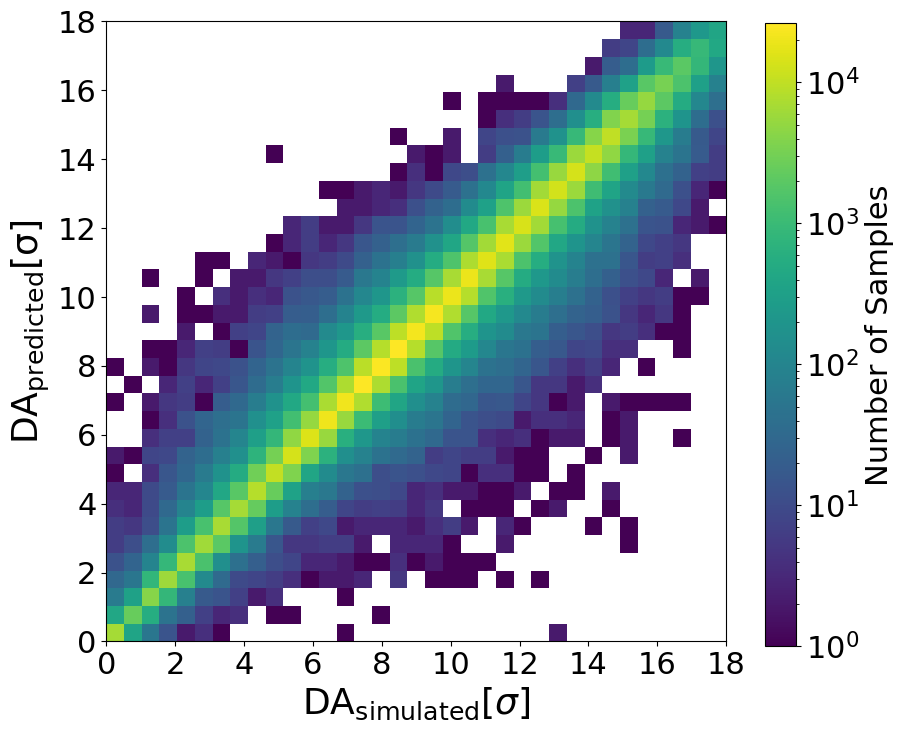}
\caption{Left: diagram of our network architecture. Right: predicted angular DA as a function of the expected angular DA values for the test data set.}
\label{fig:da_regressor}
\end{figure}

The network was developed using the TensorFlow library~\cite{tensorflow}. Architecture and hyperparameters were optimised by random search with the Keras Tuner framework~\cite{kerastuner}. 
The best model consists of four hidden layers with 2048, 1024, 512, and 128 nodes, respectively, and employs the ReLU activation function~\cite{relu}. The dropout of 1\% was implemented between the hidden layers together with the L2 regulariser in the hidden layers to improve the validation performance and avoid overfitting. Moreover, we included a concatenation layer after the MLP to explicitly incorporate the discrete variables (beam and seeds) into the model's architecture and to avoid interpolation on them. The structure of our network is shown in Fig.~\ref{fig:da_regressor} (left).

\subsection{Training and Precision}

In our initial training, we trained with 2023 data corresponding to about 15\,000 different accelerator configurations, totalling approximately 8 million samples. The loss used for the regressor is the Mean Absolute Error (MAE) function that is trained with the NADAM optimiser~\cite{nadam}. The initial learning rate is $5\times 10^{-5}$ and is halved every 10 sequential epochs if the validation loss is not improved. We found that training for 364 epochs is sufficient to validate the loss convergence. 

The MAE of the angular DA regressor is $0.34 (0.34)$ beam $\sigma$ for the test (train) data set, and the mean absolute percentage error (MAPE) is $11.91 (11.50)$, respectively. 
Furthermore, the analysis of the angular DA with the 2D histogram in Fig.~\ref{fig:da_regressor} (right) reveals that the model performs well for most of the data points, with a tight cluster around the diagonal line, indicating accurate predictions. 

\subsection{Generalising DA Prediction}

We want to ensure that the model can adapt to different optical configurations, which is a key requirement, especially in scenarios where the optics are unknown, such as in the design phase of a new accelerator. 
For this reason, we select physics variables independent of the machine optics to learn the pertinent beam dynamics features, to reduce the need for extensive retraining, and also to provide rapid adaptability to new optical configurations.
To test for this capability, we trained with data from 2023 LHC optics and tested on 2016 (non-ATS) LHC optics. Both test samples have the same size (2000 configurations each). 

\begin{figure}[!htbp]
\centering
\includegraphics[width=0.48\linewidth]{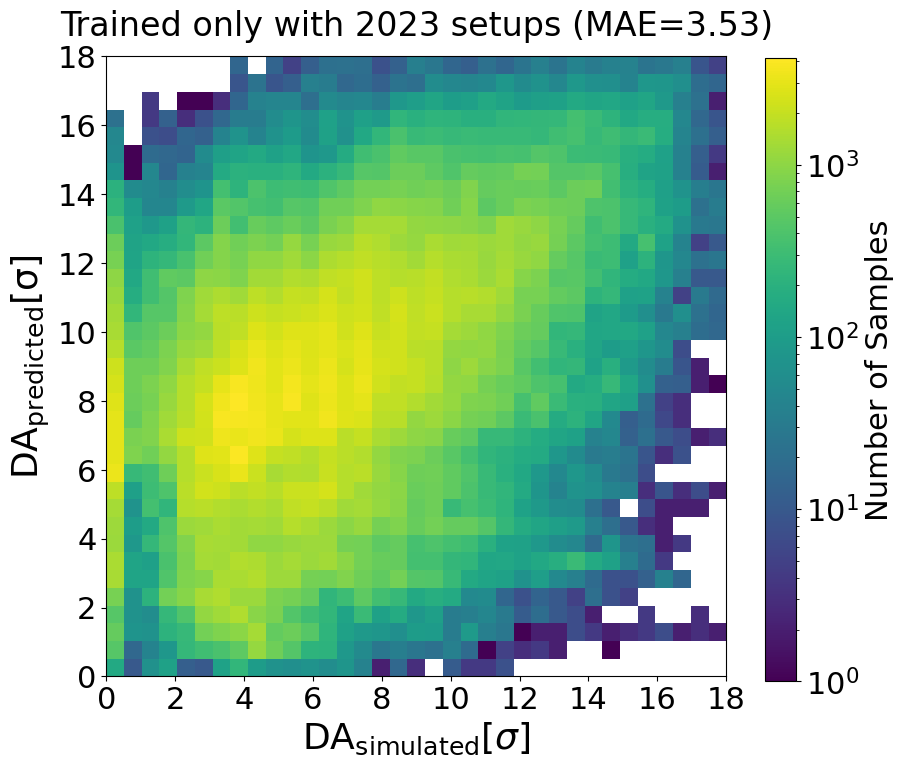}
\includegraphics[width=0.48\linewidth]{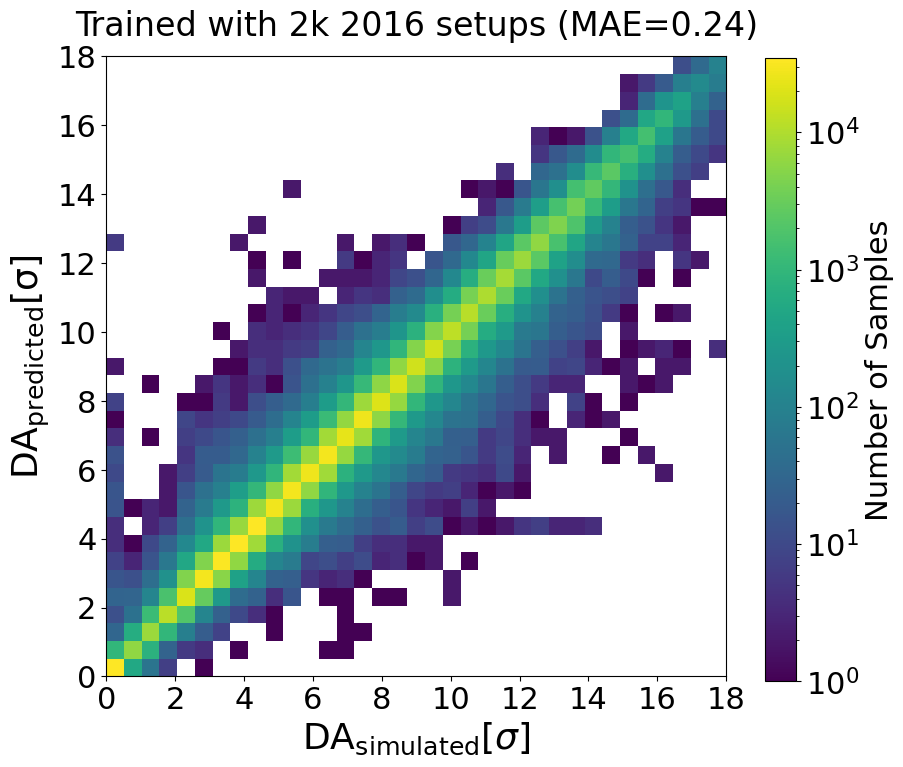}
\caption{Predicted angular DA as a function of the expected angular DA values of the 2016 test data set for the model trained only with 2023 data (left) and that trained also with 2000 configurations of the 2016 optics (right).}
\label{fig:da_2016}
\end{figure}

It turned out that the model trained with the 2023 setup is not predictive of the 2016 test data set, as indicated by its MAE = 3.53 beam $\sigma$ and the weak correlation between predicted and simulated values shown in the 2D histogram in Fig.~\ref{fig:da_2016} (left). The main reason for this is that several input variables have a different range of values for the two data sets, such as the anharmonicities and the phase advance. This causes the model to perform an inaccurate extrapolation for these variables in the 2016 data set. Unfortunately, DNNs perform well for interpolations, but, due to their non-linear nature, not for extrapolations. 
An effective approach to enhance their performance in such scenarios involves the integration of adaptive feedback mechanisms, as demonstrated in recent studies (e.g., see ~\cite{AdaptiveLearning1,AdaptiveLearning2}).
Another strategy to mitigate the extrapolation issue involves enhancing the training dataset with additional data to encompass the entire range of input variable distributions. With this intention, we have incorporated the 2016 configurations into our training dataset.

Table~\ref{tab:mae_2016} indicates the key performance figures of the model when trained with only 2023 configurations, and with the addition of 1000, 1500 or 2000 configurations from the 2016 optics, for the 2023 (500 setups) test data set, for the 2016 (500 setups) test data set, and for both test data sets together (overall).
We found out that with 2000 configurations from 2016 optics, we improved the MAE in the 2023 test to $0.33$ beam $\sigma$, but more importantly, the model became predictive in the 2016 test data set with an MAE of $0.24$ beam $\sigma$, as shown in Fig.~\ref{fig:da_2016} (right). 
\begin{table}[htbp]
\centering
\caption{Test MAE of the model trained with different amount of 2016 configurations. \label{tab:mae_2016}}
\smallskip
\begin{tabular}{c|c|c|c|c}
\hline
Data Set     & Only 2023   & +1k 2016 setups & +1.5k 2016 setups & +2k 2016 setups \\
\hline
2023 test  & 0.34 $\sigma$ & 0.37 $\sigma$ & 0.35 $\sigma$   & 0.33 $\sigma$   \\
2016 test  & 3.53 $\sigma$ & 1.64 $\sigma$ & 0.76 $\sigma$   & 0.24 $\sigma$   \\
Overall    & 1.94 $\sigma$ & 1.00 $\sigma$ & 0.54 $\sigma$   & 0.29 $\sigma$   \\
\hline
\end{tabular}
\end{table}

This ability of the model to require only a small fraction of the 2016 data to achieve predictiveness in both optics is attributed to its capacity to extract universal physics features from accelerator variables that govern the beam dynamics. This capability of rapidly adapting to new optical configurations holds immense promise for the design and optimisation of future accelerators, underscoring the critical role of surrogate models.

\subsection{Error Estimator}

We can incorporate dropout during inference, which will randomly set some of the hidden units of the network during forward propagation to zero, to exhibit variability in the angular DA prediction. The variability in predictions introduced by the dropout can subsequently serve as a basis for estimating uncertainty; this technique is known as the Monte Carlo (MC) dropout~\cite{mcdropout}. 

\begin{figure}[!htbp]
\centering
\includegraphics[height=0.44\linewidth]{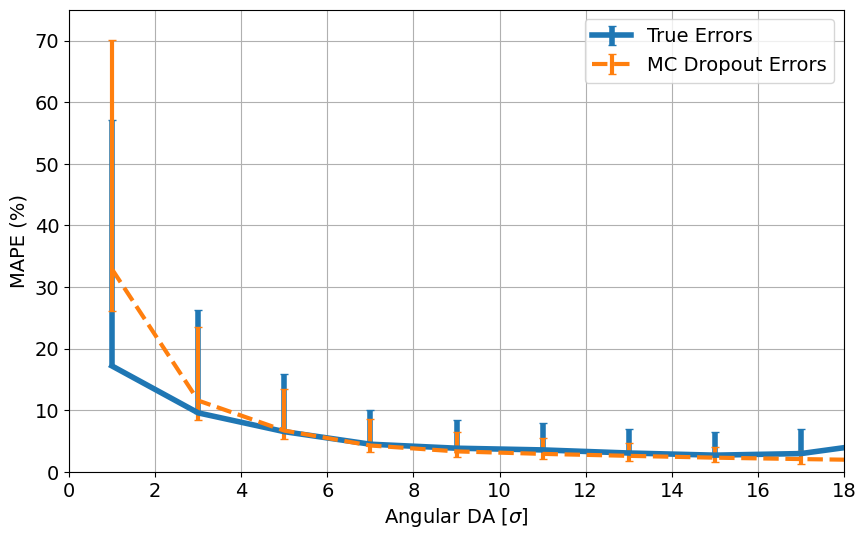}
\caption{MAPE of the true (blue) and MC dropout (orange) errors of 2023 test data set as function of the simulated angular DA.}
\label{fig:da_error}
\end{figure} 
Specifically, we introduced a 1\% dropout rate between the first two hidden layers during inference and conducted 128 inferences for each angular DA prediction. We then consider two standard deviations (std) of these predictions as the angular DA uncertainty.
The dropout rate, the number of std and the number of inferences chosen to predict the error were tuned to minimize the discrepancy between the true error (actual error in the prediction of the angular DA) and the one predicted with the MC dropout. The MAPE distributions of the true and predicted errors are shown in Fig.~\ref{fig:da_error}.

\section{Active Learning Framework}

\begin{figure}[!htbp]
\centering
\includegraphics[height=0.2\linewidth]{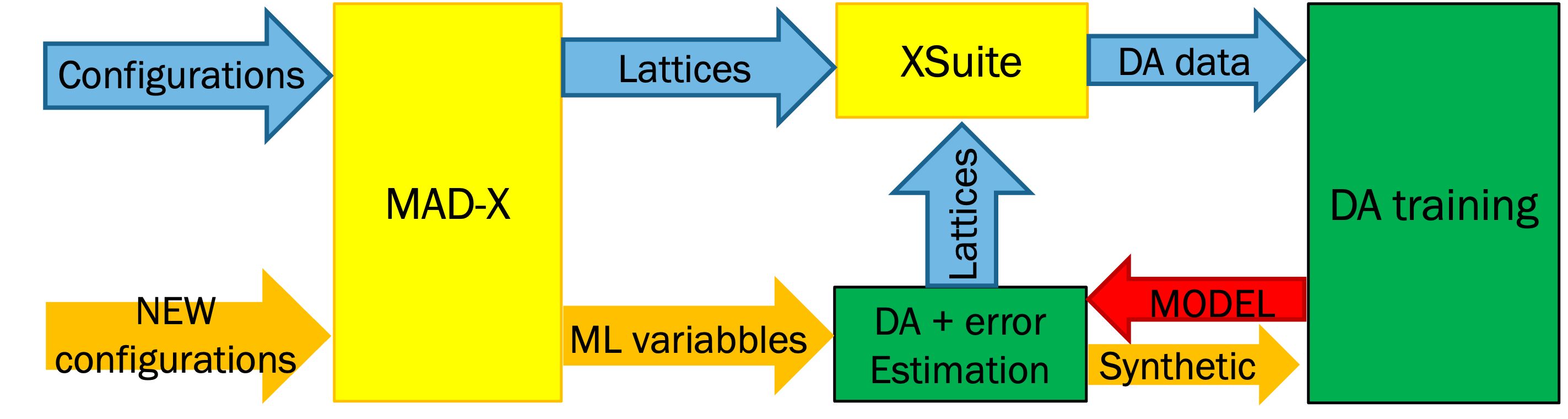}
\caption{Pipeline of the AL framework. In yellow are the full simulation/tracking algorithms and in green are the ML algorithms. The blue arrows are the full simulated data, the orange arrows are the AL data, while the red arrow indicates the angular DA model.}
\label{fig:AL_pipeline}
\end{figure}

Through the integration of the angular DA regressor and the error estimator, the AL framework becomes able to make informed sampling decisions based on the magnitude of the associated error. 
The pipeline of the AL framework is illustrated in Fig.~\ref{fig:AL_pipeline}.

\begin{figure}[!htbp]
\centering
\includegraphics[height=\linewidth]{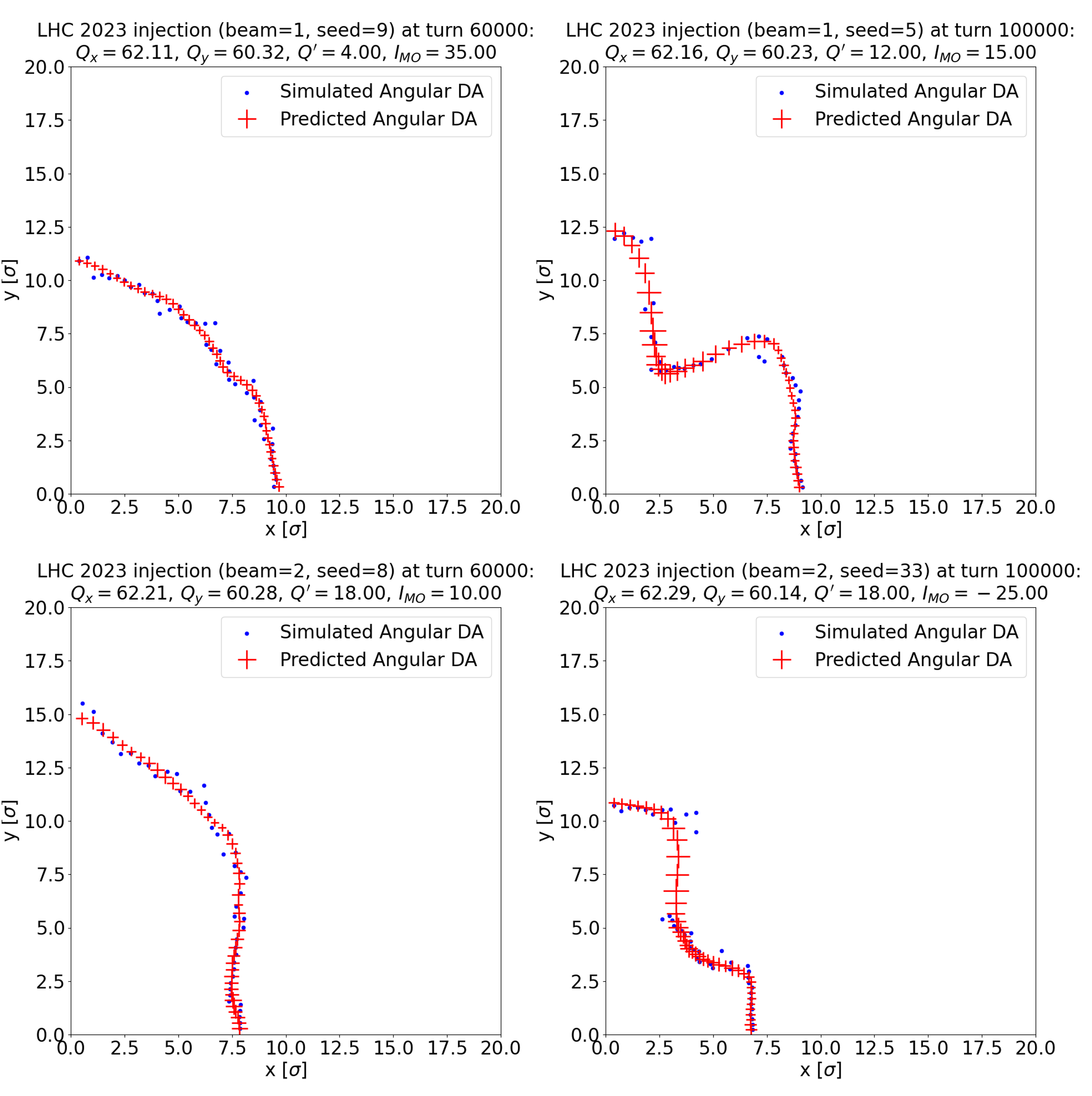}
\caption{Simulated and predicted angular DA four configurations of the test data set (predicted APE lower than 10\%).}
\label{fig:regression}
\end{figure}

In particular, when the predicted errors are small, the framework can confidently rely on the predictions of the DA regressor for a study or even to generate a synthetic data set. 
This preference is due to the impressive precision rate of 91.8\% and a negligible fallout rate of 0.1\% of our error estimator when selecting 824 configurations with Absolute Percentage Error (APE) below 10\% (which is the error divided by the angular DA) from 10\,000 accelerator configurations.
Examples of angular DA and error predictions carried out by our AL framework with APE below 10\% for four accelerator configurations (available only in the test data set) are shown in Fig.~\ref{fig:regression}.

On the contrary, when the predicted error is significant for an accelerator configuration, the system will opt to execute the entire simulation and tracking process to generate new data, allowing the ML model to learn the features of this new accelerator configuration.
Interestingly, the error estimator has an impressive accuracy rate of 99.2\% and a precision rate of 98.7\% when selecting 113 configurations with APE that exceed 200\% of the 10\,000 configurations.
By giving priority to configurations with higher predicted errors, the framework efficiently determines the order in which to simulate various accelerator configurations, a strategy referred to as smart sampling. 

\subsection{Smart Sampling}

To assess the efficacy of smart sampling, we compared the performance of the model when trained with two different data sets, both derived from the 2016 LHC optics. The first data set was generated through a random search of machine configuration input, while the second was created using our innovative active learning framework.
For this evaluation, we assembled a data set consisting of 5500 machine configurations, all obtained through random searches.

Among these, 4000 configurations (referred to as the initial training data set) were used to train the initial DA regressor, which in turn played a pivotal role in the configuration of the error estimator within the active learning framework.
Another 1000 configurations, combined with the initial 4000, were employed to assess the performance of the model trained on random search inputs (referred to as the "Random" train data set).
The remaining 500 configurations (referred to as the random testing data set) were reserved to assess the performance of the model. 

To execute the smart sampling approach, our AL framework meticulously analysed a set of 100\,000 configurations. From this extensive analysis, 1100 configurations with the highest predicted errors were selected, which were subsequently subjected to full simulation using MADX and XSuite.
Among these selected configurations, 1000 were combined with the initial data set, collectively forming what we refer to as the smart training data set with 5000 configurations. The remaining 100 configurations were reserved as a distinct test data set, labelled the smart test data set.
The performance results of the models trained with the initial, random and smart data sets are presented in Table \ref{tab:smart_sampling_performance}.
\begin{table}[htbp]
\centering
\caption{Test MAE of the model trained with different data sets. \label{tab:smart_sampling_performance}}
\smallskip
\begin{tabular}{c|c|c|c}
\hline
Data Set     & Initial train  & Random train   & Smart train \\
\hline
Random test  & 0.22 $\sigma$  & 0.21 $\sigma$  & 0.21 $\sigma$ \\
Smart test   & 1.73 $\sigma$  & 1.74 $\sigma$  & 1.70 $\sigma$ \\
\hline
\end{tabular}
\end{table}

It is worth noting that the performance of the model on the smart test data set is lower than on the random test data set. This difference arises due to the inherent complexity of the smart test data set, which translates into higher predicted errors.
In addition, the model trained with the smart training data set exhibits similar performance in the random test data set but shows an improvement in MAE when tested in the smart test data set, indicating an improvement in setups with challenging predictions.
These findings indicate the effectiveness of the smart sampling approach, which allows us to explore machine configurations more efficiently where the model has not yet learnt the features of the underlying physics. This capability may potentially shed light on the chaotic behaviour of certain parameter combinations.

\subsection{Improving Performance with Synthetic Data}

We evaluated the effectiveness of enhancing the precision of the DA Regressor with a synthetic data set generated by our AL framework. For this purpose, the framework predicted the angular DA and its associated error for 10\,000 accelerator configurations of the 2016 LHC optics, and selected 1000 configurations with the lowest predicted APE, ensuring values lower than 2\%. Variables associated with these chosen configurations, in conjunction with the predicted angular DA values, were used to construct the synthetic data set.
This synthetic data set was subsequently merged with our existing 2016 simulated data set with 5000 configurations.
The combined data set was then used to retrain our model. 
As a result, the MAE of the test data set exhibited a slight improvement, decreasing from $0.21$ beam $\sigma$ for the model trained with initial 5000 configurations to $0.20$ beam $\sigma$ for the model trained with synthetic data, compromising 6000 setups.

It is important to note the advantages of this approach in our workflow.
One significant benefit is the ability to automate learning, which allows our model to refine its prediction while waiting for the acquisition of new simulated data.
However, this improvement will inevitably reach a saturation point after a few iterations if no new simulated data is introduced. As such, despite this method offering valuable ongoing refinement, it is essential to maintain a balance between automatic learning and the incorporation of fresh data to maintain the enhancement of model performance.

In future studies, in addition to considering more simulation runs by creating synthetic data, we would like to take our existing simulation data and add some normally distributed, zero-mean random noise to the input data. This approach aims to enhance the robustness of the model without altering the angular DA values, ensuring that the information remains accurate. By introducing this noise, we can provide our model with a wider range of input variations to improve its adaptability to real-world scenarios.

\subsection{Time Performance}

The timing performance to predict the angular DA and its error for a batch of 1024 samples is \SI{140}{s} (\SI{1.41}{ms} per inference) using a Titan V GPU~\cite{titanv}, utilising 48 AMD Ryzen Threadripper 2970WX CPU cores for data loading. Taking into account a single accelerator configuration (12 bins for the survival time $\times$ 44 angles), this results in an inference of \SI{0.75}{s}/accelerator configuration.

The improvement in computational efficiency of our approach is a notable achievement. XSuite tracking, combined with the HTCondor system~\cite{htcondr_doc} which can handle around 1000 jobs concurrently, takes a waiting time about \SI{30}{h} to track particles from 1000 different configurations; equivalent to \SI{107}{s}/accelerator configuration. This figure slightly increases to \SI{147}{s}/accelerator configuration for 10\,000 configurations (roughly \SI{17}{days}).\footnote{This increase was the result of a decrease in user priority rank due to continuous use of the HTCondor system.}
Therefore, our model predictions, once trained, can be up to 200 times faster than traditional tracking methods; and with newer GPUs such as the A100~\cite{a100}, more than 2000 times faster.

It is important to note that both the AL framework and Xsuite rely on input from MAD-X, a process that takes about \SI{6.3}{\second}/accelerator configuration when executed on the HTCondor system (equivalent to approximately \SI{17}{\hour} to run 1000 MAD-X simulations in parallel). When accounting for this simulation time, which remains consistent for both the AL framework and Xsuite, the speed-up factor for the AL framework decreases to 22 times faster than traditional simulations (MAD-X+Xsuite).

\begin{table}[htbp]
\centering
\caption{Time, performance and speed-up factor to generate different size of data sets with MAD-X/Xsuite and AL framework. \label{tab:al_datasets}}
\smallskip
\begin{tabular}{c|c|c|c|c}
\hline
Data set size            & 2500          & 5000          & 10\,000           & 15\,000 \\
\hline
DA Regressor MAE          & 0.45 $\sigma$ & 0.42 $\sigma$ & 0.38 $\sigma$   & 0.34 $\sigma$   \\
Time with MAD-X + Xsuite & \SI{109}{h}         & \SI{212}{h}         & \SI{425}{h}           & \SI{612}{h}  \\
Time with AL framework   & -             & \SI{117}{h}         & \SI{126}{h}           & \SI{136}{h}  \\
Speed-up factor          & -             & 1.8           & 3.4             & 4.5   \\
\hline
\end{tabular}
\end{table}
Finally, it is important to take into account the time required for model training, which totals approximately \SI{3}{\hour} over 350 epochs.
Furthermore, we must acknowledge the need for an initial data set for training purposes.
Our findings indicate that achieving convergence requires a minimum of 2500 configurations (MAE of 0.45 for beam $\sigma$), a process that consumes approximately \SI{106}{\hour}. 
In light of this, the speed-up factor becomes more pronounced as the ratio of predicted to simulated data increases.
However, the improvement in angular DA precision is most noticeable when the initial simulated data set is substantial, as illustrated in Table~\ref{tab:al_datasets}.

\section{Conclusions}

In this paper, we present an innovative ML tool to study the beam dynamics in circular accelerators. 
Using DL networks, it was possible to accurately model and predict the DA of the CERN LHC with different optical configurations. 
One of the key takeaways of the DA regressor is the ability to extract physics-relevant features, enabling us to predict DA solely from accelerator variables. This approach not only reduces the need for extensive retraining but also provides rapid adaptability to new optical configurations, a crucial advantage for designing new accelerators where optics may be unknown.
In addition, the DA model is a powerful tool for the preservation of knowledge. By capturing and storing knowledge of physics, our models eliminate the redundancy of repetitive simulations, serving as a valuable resource for machine optimisation.

The effectiveness of knowledge preservation is significantly enhanced through the AL framework, which plays a crucial role in accelerating the learning process.
In cases where accurately predicting the DA was not possible, particularly in challenging accelerator configurations, the smart sampling strategy enabled us to first simulate these configurations. By utilising tracking tools to assess the chaotic nature of certain initial conditions, this approach significantly reduced the time required to gain insight into such features. In addition, the AL framework can generate synthetic data sets to further refine the precision of our DA model while waiting for new simulated data.
Lastly, our framework has achieved outstanding results in conducting ultrafast simulations. This remarkable speed-up proves particularly advantageous for optimising existing machines and designing future accelerators.

\acknowledgments
This work is partially funded by the SDSC project C20-10 and with the support of the Swiss Accelerator Research and Technology programme~\cite{chart}.

\bibliographystyle{JHEP}
\bibliography{biblio.bib}

\providecommand{\href}[2]{#2}\begingroup\raggedright\begin{thebibliography}{10}

\bibitem{PhysRevE.53.4067}
E.~Todesco and M.~Giovannozzi, \emph{Dynamic aperture estimates and phase-space distortions in nonlinear betatron motion}, \href{https://doi.org/10.1103/PhysRevE.53.4067}{\emph{Phys. Rev. E} {\bfseries 53} (1996) 4067}.

\bibitem{da_and_losses}
M.~Giovannozzi, \emph{A proposed scaling law for intensity evolution in hadron storage rings based on dynamic aperture variation with time}, \href{https://doi.org/10.1103/PhysRevSTAB.15.024001}{\emph{Phys. Rev. Spec. Top. Accel. Beams} {\bfseries 15} (2012) 024001}.

\bibitem{LHCDR}
O.S.~Br\"uning, P.~Collier, P.~Lebrun, S.~Myers, R.~Ostojic, J.~Poole et~al., \emph{{{LHC} Design Report}}, {CERN} Yellow Rep. Monogr., {CERN}, Geneva (2004), \href{https://doi.org/10.5170/CERN-2004-003-V-1}{10.5170/CERN-2004-003-V-1}.

\bibitem{TDR}
G.~Apollinari, I.~B\'ejar~Alonso, O.~Br\"uning, P.~Fessia, M.~Lamont, L.~Rossi et~al., \emph{{High-Luminosity Large Hadron Collider ({HL--LHC})}}, vol.~4 of \emph{{CERN} Yellow Rep. Monogr.}, {CERN}, Geneva (2017), \href{https://doi.org/10.23731/CYRM-2017-004}{10.23731/CYRM-2017-004}.

\bibitem{FCC-eeCDR}
A.~Abada et~al., \emph{{FCC}-ee: The lepton collider}, \href{https://doi.org/10.1140/epjst/e2019-900045-4}{\emph{Eur. Phys. J. Spec. Top.} {\bfseries 228} (2019) 261}.

\bibitem{FCC-hhCDR}
A.~Abada, M.~Abbrescia, S.~AbdusSalam, I.~Abdyukhanov, J.~Abelleira~Fernandez, A.~Abramov et~al., \emph{{FCC--hh: The Hadron Collider: Future Circular Collider Conceptual Design Report Volume 3. Future Circular Collider}}, \href{https://doi.org/10.1140/epjst/e2019-900087-0}{\emph{Eur. Phys. J. Spec. Top.} {\bfseries 228} (2019) 755}.

\bibitem{PhysRevE.57.3432}
M.~Giovannozzi, W.~Scandale and E.~Todesco, \emph{Dynamic aperture extrapolation in the presence of tune modulation}, \href{https://doi.org/10.1103/PhysRevE.57.3432}{\emph{Phys. Rev. E} {\bfseries 57} (1998) 3432}.

\bibitem{Bazzani:2019csk}
A.~Bazzani, M.~Giovannozzi, E.H.~Maclean, C.E.~Montanari, F.F.~Van~der Veken and W.~Van~Goethem, \emph{Advances on the modeling of the time evolution of dynamic aperture of hadron circular accelerators}, \href{https://doi.org/10.1103/PhysRevAccelBeams.22.104003}{\emph{Phys. Rev. Accel. Beams} {\bfseries 22} (2019) 104003} [\href{https://arxiv.org/abs/1909.09516}{{\ttfamily 1909.09516}}].

\bibitem{schenk:ipac21-tupab216}
M.~Schenk et~al., \emph{{Modeling Particle Stability Plots for Accelerator Optimization Using Adaptive Sampling}},  in \emph{Proc. IPAC'21}, pp.~1923--1926, JACoW Publishing, Geneva, Switzerland, 2021, \href{https://doi.org/10.18429/JACoW-IPAC2021-TUPAB216}{DOI}.

\bibitem{casanova:2023}
M.~Casanova, B.~Dalena, L.~Bonaventura and M.~Giovannozzi, \emph{Ensemble reservoir computing for dynamical systems: prediction of phase-space stable region for hadron storage rings}, \href{https://doi.org/10.1140/epjp/s13360-023-04167-y}{\emph{Eur. Phys. J. Plus} {\bfseries 138} (2023) 559}.

\bibitem{dicroce_ipac23}
D.~Di~Croce, M.~Giovannozzi, T.~Pieloni, M.~Seidel and F.F.~Van~der Veken, \emph{Accelerating dynamic aperture evaluation using deep neural networks},  in \emph{Proc. 14th Int. Particle Accelerator Conf. (IPAC'23)}, pp.~2870--2873, JACoW Publishing, Geneva, Switzerland, 2023, \href{https://doi.org/10.18429/jacow-ipac2023-wepa097}{DOI}.

\bibitem{madx}
``{MAD - Methodical Accelerator Design}.'' \url{https://mad.web.cern.ch/mad/}.

\bibitem{lhcoptics2023}
R.~Tom\'as et~al., \emph{Optics for landau damping with minimized octupolar resonances in the lhc},  in \emph{Proc. 68th {{ICFA ABDW}} on {{High-Intensity}} and {{High-Brightness Hadron Beams}}}, 2023.

\bibitem{PhysRevSTAB.16.111002}
S.~Fartoukh, \emph{Achromatic telescopic squeezing scheme and application to the lhc and its luminosity upgrade}, \href{https://doi.org/10.1103/PhysRevSTAB.16.111002}{\emph{Phys. Rev. ST Accel. Beams} {\bfseries 16} (2013) 111002}.

\bibitem{PhysRevAccelBeams.24.021002}
S.~Fartoukh, M.~Solfaroli, J.C.~De~Portugal, A.~Mereghetti, A.~Poyet and J.~Wenninger, \emph{Achromatic telescopic squeezing scheme and by-products: From concept to validation}, \href{https://doi.org/10.1103/PhysRevAccelBeams.24.021002}{\emph{Phys. Rev. Accel. Beams} {\bfseries 24} (2021) 021002}.

\bibitem{xsuite_doc}
``Xsuite documentation.'' http://xsuite.web.cern.ch, 2023.

\bibitem{xsuite_gianni}
G.~Iadarola et~al., \emph{Xsuite: an integrated beam physics simulation framework},  in \emph{Proc. 68th {{ICFA ABDW}} on {{High-Intensity}} and {{High-Brightness Hadron Beams}}}, 2023.

\bibitem{ptc_doc}
``{MAD-X-PTC documentation}.'' http://mad.web.cern.ch/mad/madX/doc/usrguide/ptc\_general/ptc\_general.html, 2006.

\bibitem{cook}
R.D.~Cook and S.~Weisberg, \emph{Influential observations, high leverage points, and outliers in linear regression}, {\emph{J. Am. Stat. Assoc.} {\bfseries 74} (1982) 169}.

\bibitem{lecun}
Y.~LeCun, Y.~Bengio and G.~Hinton, \emph{Deep learning}, {\emph{Nature} {\bfseries 521} (2020) 436}.

\bibitem{rnn_gradient}
Y.~Bengio, P.~Simard and P.~Frasconi, \emph{Learning long-term dependencies with gradient descent is difficult}, \href{https://doi.org/10.1109/72.279181}{\emph{IEEE Transactions on Neural Networks} {\bfseries 5} (1994) 157}.

\bibitem{imagenet}
A.~Krizhevsky, I.~Sutskever and G.E.~Hinton, \emph{{ImageNet Classification with Deep Convolutional Neural Networks}},  in \emph{Advances in Neural Information Processing Systems}, F.~Pereira, C.~Burges, L.~Bottou and K.~Weinberger, eds., vol.~25, Curran Associates, Inc., 2012.

\bibitem{tensorflow}
M.~Abadi et~al., \emph{Tensorflow: Large-scale machine learning on heterogeneous systems}, .

\bibitem{kerastuner}
A.O.~Gomes et~al., ``Keras tuner.'' GitHub repository.

\bibitem{relu}
V.~Nair and G.E.~Hinton, \emph{Rectified linear units improve restricted boltzmann machines vinod nair},  in \emph{Proceedings of ICML}, vol.~27, pp.~807--814, 06, 2010.

\bibitem{nadam}
T.~Dozat, \emph{Incorporating {Nesterov} momentum into {Adam}},  in \emph{Proc. 4th Int. Conf. on Learning Representations (ICLR'16)}, 2016.

\bibitem{AdaptiveLearning1}
A.~Scheinker, A.~Edelen, D.~Bohler, C.~Emma and A.~Lutman, \emph{Demonstration of model-independent control of the longitudinal phase space of electron beams in the linac-coherent light source with femtosecond resolution}, \href{https://doi.org/10.1103/PhysRevLett.121.044801}{\emph{Phys. Rev. Lett.} {\bfseries 121} (2018) 044801}.

\bibitem{AdaptiveLearning2}
A.~Scheinker, F.~Cropp and D.~Filippetto, \emph{Adaptive autoencoder latent space tuning for more robust machine learning beyond the training set for six-dimensional phase space diagnostics of a time-varying ultrafast electron-diffraction compact accelerator}, \href{https://doi.org/10.1103/PhysRevE.107.045302}{\emph{Phys. Rev. E} {\bfseries 107} (2023) 045302}.

\bibitem{mcdropout}
Y.~Gal and Z.~Ghahramani, \emph{Dropout as a bayesian approximation: Representing model uncertainty in deep learning},  in \emph{Proceedings of the 33rd International Conference on International Conference on Machine Learning - Volume 48}, ICML'16, pp.~1050--1059, JMLR.org, 2016, \href{http://jmlr.org/proceedings/papers/v48/gal16.html}{http://jmlr.org/proceedings/papers/v48/gal16.html}.

\bibitem{titanv}
{NVIDIA Corporation}, \emph{{NVIDIA TITAN V}},  2017.

\bibitem{htcondr_doc}
``{CERN HT-Condor documentation}.'' https://batchdocs.web.cern.ch/local/submit.html, 2023.

\bibitem{a100}
{NVIDIA Corporation}, \emph{{NVIDIA A100}},  2020.

\bibitem{chart}
``{Accelerator Research and Technology programme - CHART}.'' \url{http://www.chart.ch}.

\end{thebibliography}\endgroup

\end{document}